\documentclass[aps,12pt,pra,showpacs,superscriptaddress,preprint]{revtex4}

\usepackage{amsmath}

\usepackage[dvipdfmx]{graphicx}

\usepackage{subfig}

\usepackage{array}

\usepackage{amsfonts}

\usepackage{epsf}

\usepackage{epsfig}

\usepackage{amssymb}

\usepackage{bbm}

\usepackage{delarray}

\usepackage{caption}

\usepackage{amstext}

\usepackage{graphics}

\catcode`ð=\active

\defð{\u{g}}

\catcode`Ð=\active

\defÐ{\u{G}}

\catcode`Ý=\active

\defÝ{\. I}

\catcode`ö=\active

\defö{\"{o}}

\catcode`Ö=\active

\defÖ{\"O}

\catcode`ü=\active

\defü{\"{u}}

\catcode`Ü=\active

\defÜ{\"{U}}

\catcode`Þ=\active

\defÞ{\c{S}}

\catcode`þ=\active

\defþ{\c{s}}

\catcode`ý=\active

\defý{{\i}}

\catcode`ç=\active

\defç{\d{c}}

\catcode`Ç=\active

\defÇ{\d{C}}

\begin{document}

\title{Effective Mass Quantum Systems with Displacement Operator: Inverse Square plus Coulomb-like Potential}
\author{\small Altuð Arda}
\email[E-mail: ]{arda@hacettepe.edu.tr}\affiliation{Department of
Physics Education, Hacettepe University, 06800, Ankara,Turkey}
\author{\small Ramazan Sever}
\email[E-mail: ]{sever@metu.edu.tr}\affiliation{Department of
Physics, Middle East Technical  University, 06531, Ankara,Turkey}

\begin{abstract}
The Schrödinger-like equation written in terms of the displacement
operator is solved analytically for a inverse square plus Coulomb-like potential.
Starting from the new Hamiltonian, the effects of the spatially dependent mass on the bound states
and normalized wave functions
of the "usual" inverse square plus Coulomb interaction are discussed.
\end{abstract}


\maketitle

\newpage

\section{Introduction}
It has been argued that the divergences appearing in field
theories can be removed with the idea of noncommutativity and this
can be done by using a universal invariant length parameter. This
approach has became a central idea in the physical and
mathematical points of view [1, 2]. Within the quantum mechanics,
the noncommutative coordinates written by the terms of minimum
length scale lead to some modifications in position-momentum
commutators [3, 4]. This yields an extended Hamiltonian having a
position-dependent mass term in kinetic part with some ambiguity
parameters $\alpha, \beta, \delta$ satisfying
$\alpha+\beta+\delta=1$, and a Schrödinger equation with effective
mass [5-7].

Recently, Filho and co-workers have analysed a quantum system with
position-dependent mass by using a different approach where they
suggest a displacement operator given by
\begin{eqnarray}
T_{\gamma}(dx)|x>=|x+dx+\gamma xdx>\,,
\end{eqnarray}
where $\gamma$ is a real constant describing the mixing between
the displacement and the original position state. This operator
transforms a well-localized state around $x$ to another
well-localized state around $x+(1+\gamma x)dx$ while all other
physical properties remain unchanged [8]. This operator is written
explicitly as
\begin{eqnarray}
T_{\gamma}(dx)=I-\,\frac{i}{\hbar}\,\hat{p}_{\gamma}dx\,,
\end{eqnarray}
where $\hat{p}_{\gamma}$ corresponds to the generalized linear
momentum operator. The commutator between $\hat{p}_{\gamma}$ and
$\hat{x}$ operator is written as by
$[\hat{x},\hat{p}_{\gamma}]=(1+\gamma x)i\hbar$ which gives a
generalized uncertainty relation
\begin{eqnarray}
\Delta x\Delta p_{\gamma} \geq (1+\gamma<x>)\,\frac{\hbar}{2}\,,
\end{eqnarray}

The generalized momentum operator can be given as [8]
\begin{eqnarray}
\hat{p}_{\gamma}|\alpha>=-i\hbar(1+\gamma
x)\frac{d}{dx}|\alpha>\,,
\end{eqnarray}
and the corresponding deformed derivative is written as
$D_{\gamma}=(1+\gamma x)\,\frac{d}{dx}$ where
$\hat{p}_{\gamma}=-i\hbar D_{\gamma}$. In Ref. [9], the
generalized momentum operator is written in Hermitian form which
enables us to write the Hamiltonian of the system as a Hermitian
operator. 

The time-dependent form of the equation for a particle moving in a potential field $V(x)$ is written as stated by Filho and co-workers [10]  
\begin{eqnarray}
\left\{i\hbar\,\frac{\partial}{\partial t}+\frac{\hbar^2}{2m}(1+\gamma x)\left[(1+\gamma x)\frac{\partial^2}{\partial x^2}+\gamma\frac{\partial}{\partial x}\right]-V(x)\right\}\Psi(x,t)=0\,,
\end{eqnarray}
If we consider the time-independent Hamiltonian operator to be
$H=\hat{p}^2_{\gamma}/2m+V(x)$, we assume the wave function as $\Psi(x,t)=\phi(x)e^{-iEt/\hbar}$, and obtain the following
Schrödinger-like equation for a single particle  
\begin{eqnarray}
\left[-\frac{\hbar^2}{2m}D^2_{\gamma}-E+V(x)\right]\phi(x)=0\,,
\end{eqnarray}
or
\begin{eqnarray}
\left[\frac{2}{m(x)}\,\frac{d^2}{dx^2}+\,\frac{d}{dx}\left(\frac{1}{m(x)}\right)\,\frac{d}{dx}+\frac{4}{\hbar^2}[E-V(x)]\right]\phi(x)=0\,,
\end{eqnarray}
with $m(x)=m(1+\gamma x)^{-2}$. In Refs. [8, 9], the authors have
tested their ideas for a free particle, and a particle moving
in a one-dimensional infinite well of length $L$. They have 
obtained analytical solutions for the above systems. They have discussed the expectation values of the position, and the normalization of the wave functions. In Ref. [8], the authors have also studied the
dependence of the transmission and tunnelling probability on
parameter $\gamma$ for a particle subjected to a potential barrier
with height $V_{0}>0$. In Ref. [11], the general bound state solutions and the corresponding normalized
wave functions have been discussed for a potential function including a quartic and a
quadratic term. In the present work, starting from the
Schrödinger-like equation given in Eq. (7), we will search the
analytical solutions for a particle moving in an inverse square plus Coulomb-like
potential of the form
\begin{eqnarray}
V(x)=\frac{A}{x^2}-\frac{B}{x}\,,\nonumber
\end{eqnarray}
Our aim is to find the bound states
and to see the effect of the parameter $\gamma$ on the energy
eigenvalues. We will also find the wave functions with their
normalization constants.

\section{Analytical Solutions}
In order to study the effects of displacament operator on the results of the present
problem, we change the variable to $z=1+\gamma x$, and write the
above potential function into Eq. (7) giving
\begin{eqnarray}
\frac{d^2\phi(z)}{dz^2}+\frac{1}{z}\frac{d\phi(z)}{dz}+\left[\frac{a_1}{z^2}+\frac{a_2}{z(1-z)}+\frac{a_3}{(1-z)^2}\right]\phi(z)=0\,,
\end{eqnarray}
where
\begin{subequations}
\begin{align}
a_1&=M[E-\gamma(A\gamma+B)]\,,\\
a_2&=-\gamma M(2A\gamma+B)\,,\\
a_3&=-AM\gamma^2\,,
\end{align}
\end{subequations}
with $M=2m/\gamma^2\hbar^2$.

The transformation on the wave function such as $\phi(z)=z^{p}(1-z)^{q}\psi(z)$ gives a second order differential equation
\begin{eqnarray}
z(1-z)\frac{d^2\psi(z)}{dz^2}+[1+2p-(1+2p+2q)z]\frac{d\psi(z)}{dz}+(a_2-2pq-q)\psi(z)=0\,,
\end{eqnarray}
which could be a hypergeometric differential equation if the parameters used in the equation satisfy [12-14]
\begin{eqnarray}
p^2=M[\gamma(A\gamma+B)-E]\,\,\,;\,\,q=\frac{1}{2}\,\left[1\pm\sqrt{1+4AM\gamma^2\,}\right]\,,
\end{eqnarray}
Comparing with the hypergeometric differential equation as following
\begin{eqnarray}
y(1-y)\frac{d^2\omega}{dy^2}+[c-(a+b+1)y]\frac{d\omega}{dy}-ab\omega=0\,,
\end{eqnarray}
we obtain the solution of Eq. (9) as
\begin{eqnarray}
\psi(z) \sim\,_{2}F_{1}(a,b;c;z)=\sum_{k=0}^{\infty}\frac{(a)_{k}(b)_{k}}{(c)_{k}}\,\frac{z^k}{k!}\,,
\end{eqnarray}
where $\,_{2}F_{1}(a,b;c;z)$ is the hypergeometric function and $(a)_{k}$ is Pochhammer symbol [12-14]. The new parameters in $\,_{2}F_{1}(a,b;c;z)$ should be
\begin{eqnarray}
a=p+q-i\sqrt{ME\,}\,\,;\,\,b=p+q+i\sqrt{ME\,}\,\,;\,\,c=1+2\sqrt{M[\gamma(A\gamma+B)-E]\,}\,,
\end{eqnarray}

The mathematical solutions of Eq. (8) is written as
\begin{eqnarray}
\phi(z)=Nz^{p}(1-z)^{q}\,_{2}F_{1}(a,b;c;z)\,.
\end{eqnarray}
where the normalization constant $N$ is obtained below. We now obtain the energy eigenvalues of the system in the next section.
\subsection{Energy Spectrum}

In order to obtain a physical solution for the wave functions, the parameter $a$ in $\,_{2}F_{1}(a,b;c;z)$ should be $a=-n\,\,(n=0, 1,2,\ldots)$ which is the quantization rule of the system and gives us the energy eigenvalues as
\begin{eqnarray}
E(n,\gamma,A,B)=-\frac{1}{4}\left[\frac{\gamma\hbar}{\sqrt{2m\,}}
\,\left(n+\frac{1}{2}+\frac{1}{2}\sqrt{1+\frac{8Am}{\hbar^2}\,}\right)
-\frac{\sqrt{2m\,}}{\hbar}\,\frac{A\gamma+B}{n+\frac{1}{2}+\frac{1}{2}\sqrt{1+\frac{8Am}{\hbar^2}\,}}\right]^2\,,\nonumber\\
\end{eqnarray}
Firstly, we want to compare our results analytically for the case of "constant" mass. To achieve this aim, we introduce the principal quantum number as $N=n+1$. In this case, the energy eigenvalues are written as
\begin{eqnarray}
E(N,\gamma,A,B)=-\frac{1}{4}\left[\frac{\gamma\hbar}{2\sqrt{2m\,}}
\,\left(2N-1+\sqrt{1+\frac{8Am}{\hbar^2}\,}\right)
-\frac{2\sqrt{2m\,}}{\hbar}\,\frac{A\gamma+B}{2N-1+\sqrt{1+\frac{8Am}{\hbar^2}\,}}\right]^2\,,\nonumber\\
\end{eqnarray}
We obtain the following from the last equation
\begin{eqnarray}
E(N,0,A,B)=-\frac{1}{4}\left[\frac{2\sqrt{2m\,}}{\hbar}\,\frac{B}{2N-1+\sqrt{1+\frac{8Am}{\hbar^2}\,}}\right]^2\,,
\end{eqnarray}
which is exactly the same with Eq. (15) given in Ref. [15]. Eq. (17) gives the result for "usual" Coulomb-like potential for the case of "constant" mass
\begin{eqnarray}
E(N,0,0,B)=-\frac{2m}{\hbar^2}\,\frac{B^2}{4N^2}\,,
\end{eqnarray}
which is the same with Eq. (27) obtained in Ref. [15]. It is suitable now to give the result for the Coulomb-like potential for the case where the mass depends on spatially coordinate
\begin{eqnarray}
E(N,\gamma,0,B)=\frac{1}{2}\gamma B-\frac{\gamma^2\hbar^2}{32m}n'^2-\frac{2m}{\hbar^2}\frac{B^2}{n'^2}\,.
\end{eqnarray}
with $n'=2N-1$.

Secondly, we summarize the numerical results in Table I. In general, the numerical analyse for such potentials are computed for diatomic molecules. So, we give the bound state energies for a diatomic molecule ($CO$ molecule) in $eV$ where the parameter values used here are taken from Ref. [16] inserting a new quantity $E_{0}=\hbar^2/mr_{0}^2$ and, by comparing, we set the potential parameters as $A=D_{e}r_{e}^2$ and $B=2D_{e}r_{e}$ ($D_{e}$ is the dissociation energy and $r_{e}$ is the equilibrium distance). By using Eq. (16), the binding energies are calculated four different values of parameter $\gamma$ to see the effect of position-dependent mass on energy levels of the inverse square plus Coulomb-like potential. It is also seen in Table I that we give some numerical results for the potential for the constant mass case ($\gamma=0$). There is an increasingly contribution of the parameter $\gamma$ on the energy levels of the potential. Finally, we summarize some numerical energy eigenvalues obtaining from Eq. (20) for the Coulomb-like potential ($A=0$) for both two cases of $\gamma \neq 0$ and $\gamma=0$, respectively, in Table II. Here, our aim is just to giving an idea about the effect of parameter $\gamma$ on the energy levels for the Coulomb problem, the results are obtained in atomic units. It is seen that the contribution of parameter $\gamma$ increases while it's value increases. In Tables, we use only the values for $\gamma$ falling into the range $[0,1]$. The dependency of the energy eigenvalues on the displacement operator for the case where the $\gamma$-values greater than one are given in Figure I. We plot the variation of energy eigenvalue only for ground states for the inverse square plus Coulomb-like potential, and Coulomb-like potential, respectively, because the shape for the upper energy levels are similar. It is observed that the results in Fig. I are consistent with the ones given in Tables. 
\subsection{Normalization}
The wave functions should be satisfy
\begin{eqnarray}
\int_{-\infty}^{+\infty}|\phi(z)|^2dz=1\,,
\end{eqnarray}
By using the following identity for the hypergeometric functions for $|z|\rightarrow \infty$ [14]
\begin{eqnarray}
\,_{2}F_{1}(a,b;c;z)=\frac{\Gamma(b-a)\Gamma(c)}{\Gamma(b)\Gamma(c-a)}\,(-z)^{-a}+\frac{\Gamma(a-b)\Gamma(c)}{\Gamma(a)\Gamma(c-b)}\,(-z)^{-b}\,,
\end{eqnarray}
Eq. (21) is written as
\begin{eqnarray}
2|N|^2\int_{0}^{\infty}(z)^{2p}(1-z)^{2q}\left(\Gamma_{1}z^{n}+\Gamma_{2}z^{-(n+2p+2q)}\right)^2dz=1\,,
\end{eqnarray}
where
\begin{eqnarray}
\Gamma_{1}=\,(-1)^{n}\,\frac{\Gamma(2n+2p+2q)\Gamma(1+2p)}{\Gamma(n+2p+2q)\Gamma(1+n+2p)}\,\,;\,\,
\Gamma_{2}=\,(-1)^{-(n+2p+2q)}\,\frac{\Gamma(-2n-2p-2q)\Gamma(1+2p)}{\Gamma(-n)\Gamma(1-n-2q)}\,,\nonumber\\
\end{eqnarray}
By defining a new variable such as $y=z/(1+z)$ we can use the integral equation [13, 14]
\begin{eqnarray}
\int_{0}^{1}t^{r-1}(1-t)^{r'-1}(1-tx)^{-r-r'}dt=B(r,r')\,_{2}F_{1}(r+r',r;r+r';x)\,,
\end{eqnarray}
where $B(a',b')$ is the Beta integral [12-14]. Using Eq. (25) gives us the normalization constant as
\begin{eqnarray}
N=\frac{1}{\sqrt{2\,}}\,\frac{1}{\sqrt{I_{1}+I_{2}+I_{3}\,}}\,,
\end{eqnarray}
where
\begin{eqnarray}
I_{1}&=&\Gamma_{1}^{2}B(1+2n+2p,-1-2n-2p-2q)\,_{2}F_{1}(-2q,1+2n+2p;-2q;2)\,,\nonumber\\
I_{2}&=&2\Gamma_{1}\Gamma_{2}B(1-2q,-1)\,_{2}F_{1}(-2q,1-2q;-2q;2)\,,\nonumber\\
I_{3}&=&\Gamma_{2}^{2}B(1-2n-2p-4q,-1-2n-2p-2q)\,_{2}F_{1}(-2q,1-2n-2p-2q;-2q;2)\,.
\end{eqnarray}

\section{Conclusion}
Starting from the Schrödinger-like equation written in terms of
the translation operator, we have analysed the changes of the
bound states of a inverse square plus Coulomb-like potential. We have computed the
corresponding normalized wave functions analytically. We have also
given two tables, and a figure to see the variation of the bound states according to
the parameter $\gamma$. We have found that our analytical results obtained
for the bound states are in agreement with the ones obtained for
the case where the mass is constant as $\gamma \rightarrow 0$.

\section{Acknowledgments}
One of authors (A.A.) thanks Prof Dr Andreas Fring from City University London and the Department of Mathematics for hospitality where last part of the manuscript has been completed. This research was partially supported by the Scientific and
Technical Research Council of Turkey.

\newpage

\begin{table}
\begin{ruledtabular}
\caption{Energy eigenvalues for the inverse square plus Coulomb-like potential.}
\begin{tabular}{ccccc}
$n$ & $\gamma=0$ & $\gamma=0.1$ & $\gamma=0.5$ & $\gamma=1.0$ \\
\hline
0 & 0.051710 & 0.058521 & 0.082237 & 0.111846  \\
1 & 0.153947 & 0.172279 & 0.241986 & 0.328809  \\
2 & 0.254787 & 0.284476 & 0.399400 & 0.542203  \\
3 & 0.354256 & 0.395141 & 0.554523 & 0.752090  \\
4 & 0.452378 & 0.504302 & 0.707395 & 0.958530  \\
5 & 0.549178 & 0.611985 & 0.858054 & 0.958530  \\
\end{tabular}
\end{ruledtabular}
\end{table}

\begin{table}
\begin{ruledtabular}
\caption{Energy eigenvalues ($-E$) for the Coulomb-like potential ($B=5$).}
\begin{tabular}{ccccc}
$n$ & $\gamma=0$ & $\gamma=0.1$ & $\gamma=0.5$ & $\gamma=1.0$ \\
\hline
0 & 12.5000 & 12.2513 & 11.2813 & 10.1250  \\
1 & 3.12500 & 2.88000 & 2.00000 & 1.12500  \\
2 & 1.38889 & 1.15014 & 0.42014 & 0.01389  \\
3 & 0.78125 & 0.55125 & 0.03125 & 0.28125  \\
4 & 0.50000 & 0.28125 & 0.03125 & ---  \\
5 & 0.34722 & 0.14222 & 0.22222 & ---  \\
\end{tabular}
\end{ruledtabular}
\end{table}

\newpage 

\begin{figure}
\centering 
\subfloat[][energy for inverse square plus Coulomb-like potential.]{\includegraphics[height=2in,
width=3in, angle=0]{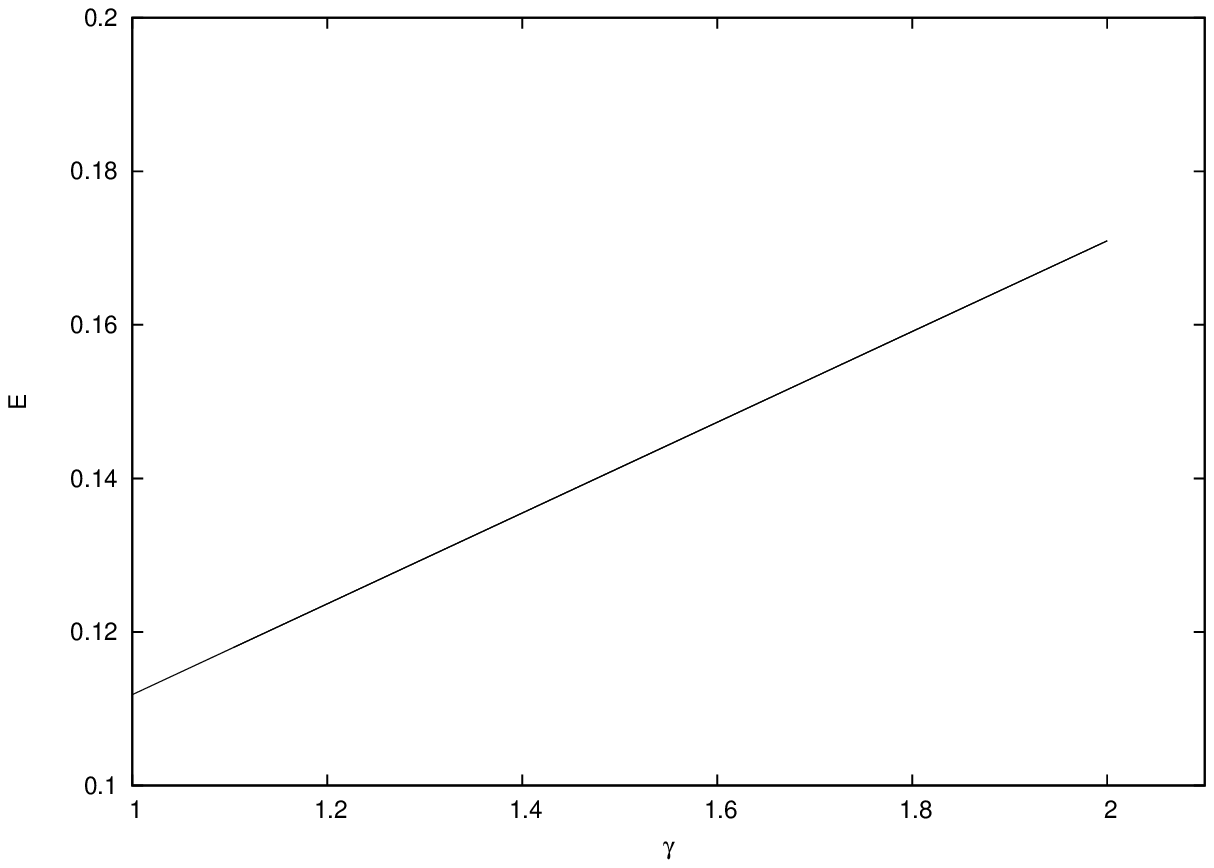}}
\subfloat[][energy for Coulomb-like potential ($-E$).]{\includegraphics[height=2in, width=3in,
angle=0]{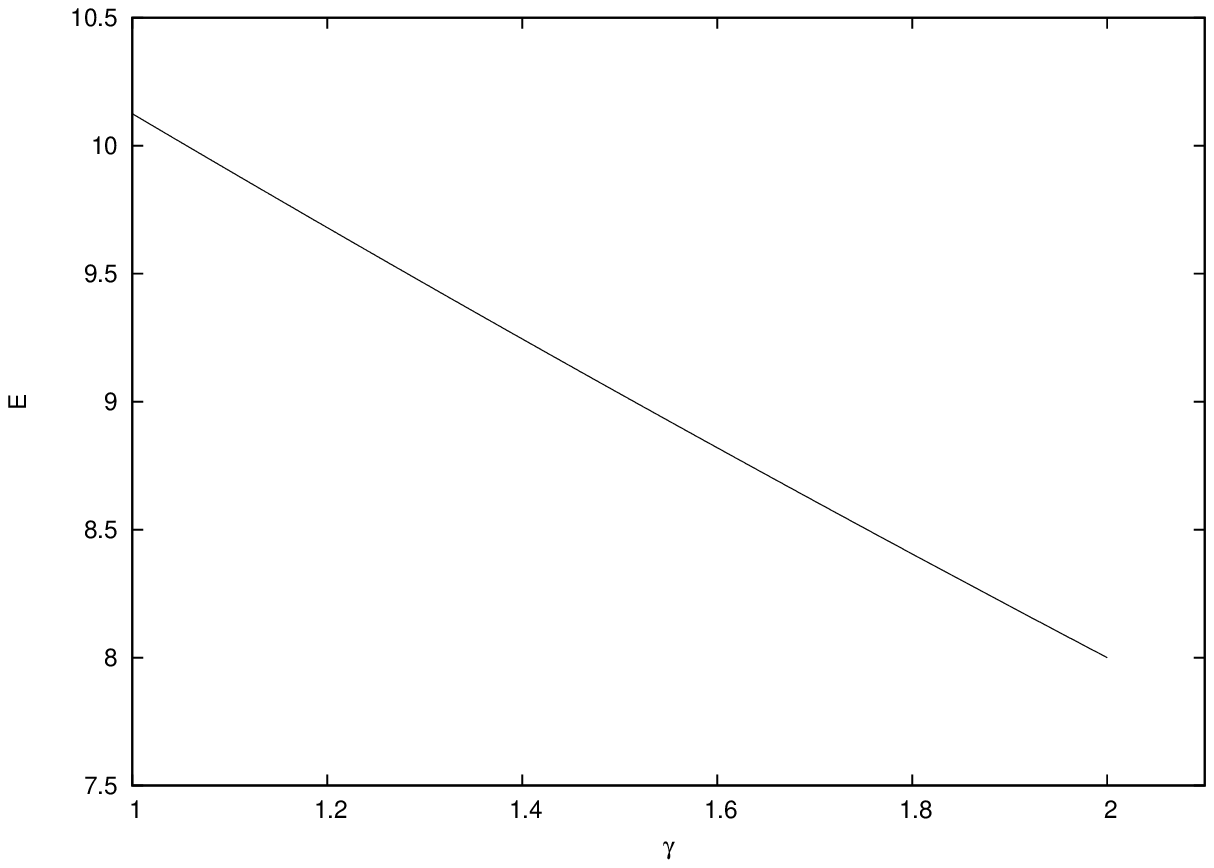}}
 \caption{The dependencies of energy eigenvalues for the inverse square plus Coulomb-like potential (left panel), and the Coulomb-like potential (right panel) on mixing parameter $\gamma$.}
\end{figure}


\newpage



\begin{thebibliography}{99}

\bibitem{ref1} H. S. Synder, Phys. Rev. {\bf 71} (1947) 38.

\bibitem{ref2} A. Connes, \textit{Noncommutative Geometry}, Academic Press (1994).

\bibitem{ref3} A. Kempf, G. Mangano, and R. B. Mann, Phys. Rev. D {\bf 52} (1995) 1108.

\bibitem{ref4} J. Gamboa, M. Loewe, and J. C. Rojas, Phys. Rev. D {\bf 64} (2001) 067901.

\bibitem{ref5} O. von Roos, Phys. Rev. B {\bf 27} (1983) 7547.

\bibitem{ref6} D. J. BenDaniel, and C. B. Duke, Phys. Rev. {\bf 152} (1966) 683.

\bibitem{ref7} A. S. Dutra, and C. A. S. Almeida, Phys. Lett. A {\bf 275} (2000) 25.

\bibitem{ref8} R. N. Costa Filho, M. P. Almeida, G. A. Farias, and J. S. Andrade Jr, Phys. Rev. A {\bf 84} (2011) 050120.

\bibitem{ref9} S. H. Mazharimousavi, Phys. Rev. A {\bf 85} (2012) 034102.

\bibitem{ref10} R. N. Costa Filho, G. Alencar, B. Skagerstam, and J. S. Andrade Jr, EPL, {\bf 101} (2013) 10009.

\bibitem{ref11} M. Vubangsi, M. Tchoffo, and L. C. Fai, Phys. Scr.
{\bf 89} (2014) 025101.

\bibitem{ref12} C. W. Wong, \textit{Introduction to Mathematical Physics-Methods
and Concepts }, Oxford University Press, New York (1991).

\bibitem{ref13} M. Abramowitz, and I. A. Stegun (Eds.), \textit{Handbook of Mathematical
Functions with Formulas, Graphs, and Mathematical Tables}, Dover Publications, New York (1965).

\bibitem{ref14} I. S. Gradshteyn, and I. M.Ryzhik, \textit{Table of Integrals, Series and Products}, eds. A. Jeffrey, and D. Zwillinger,  Academic Press, New York (2007).

\bibitem{ref15} S. H. Dong, and G. H. Sun, Phys. Scr. {\bf 70} (2004) 94.

\bibitem{ref16} I. Nasser, M. S. Abdelmonem, H. Bahlouli, and A. D. Alhaidari, J. Phys. B {\bf 40} (2007) 4245.
\end{thebibliography}
\end{document}